\documentclass[conference]{IEEEtran}
\usepackage[latin1]{inputenc}
\usepackage[english]{babel}
\usepackage{amsbsy}
\usepackage{amsmath}
\usepackage{amssymb}
\usepackage{bbding}
\usepackage{empheq}
\usepackage{latexsym}
\usepackage[dvips]{graphicx}
\usepackage{textcomp}
\usepackage{relsize}

\begin{document}
\title{Impact of Scheduling in the Return-Link of Multi-Beam Satellite MIMO Systems}

\author{
Vincent~Boussemart,~Loris~Marini,~Matteo~Berioli\\
Institute of Communications and Navigation, German Aerospace Center (DLR)\\Oberpfaffenhofen, 82234 We\ss ling, Germany\\
E-mail: {\tt \{vincent.boussemart, loris.marini, matteo.berioli\}@dlr.de}
}

\maketitle

\begin{abstract}
The utilization of universal frequency reuse in multi-beam satellite systems introduces a non-negligible level of co-channel interference (CCI), which in turn penalizes the quality of service experienced by users. Taking this as starting point, the paper focuses on resource management performed by the gateway (hub) on the return-link, with particular emphasis on a scheduling algorithm based on bipartite graph approach. The study gives important insights into the achievable per-user rate and the role played by the number of users and spot beams considered for scheduling. More interestingly, it is shown that a free-slot assignment strategy helps to exploit the available satellite resources, thus guaranteeing a max-min rate requirement to users. Remarks about the trade-off between efficiency-loss and performance increase are finally drawn at the end of the paper.
\end{abstract}

\IEEEpeerreviewmaketitle

\section{Introduction}\label{sec:Introduction}

Current satellites are built with hundreds of spot beams in order to extend the coverage area and also to increase the antenna gains \cite{MaBo02}. The frequency allocated to each spot beam is not necessarily the same and actually depends on the number of frequencies present in the system, also called number of colors \cite{LuWeJa00}. It is known that the utilization of universal frequency reuse can lead to an increase of capacity at the expense of co-channel interference which becomes rather critical.

Employing interference cancellation (IC) techniques in the return-link, it is possible to diminish the co-channel interference (CCI) caused by the users on ground. These operations can be performed by the gateway since it receives all user's signals. In this way it was shown in \cite{BBRJ11} that the maximum achievable rate in the return-link of a multi-beam satellite system is higher for universal frequency reuse than for higher numbers of colors.

From information theory, the maximum achievable rate of a multiple-input multiple-output (MIMO) system can be computed using Telatar formula \cite{Tel99}. It is known that a receiver implementing Minimum Mean Square Error (MMSE) with optimal successive interference cancellation (SIC) architecture is capacity achieving when having perfect channel state information (CSI) \cite{TsVi08}. The order in which the SIC operation is performed modifies the achievable per-user  rates and maximizing the minimum rate can be realized using the Foschini algorithm \cite{FGVW99} with MMSE filtering as in \cite{BBRJ11}.

According to the current standard for the return satellite link \cite{DVB-RCS} CCI will occur for instance when users are assigned to the same multi-frequency - time division multiple access (MF-TDMA) slot in different beams. Since in clear sky conditions CCI mainly depends on the users' position on ground, it is possible to properly choose the set of the users transmitting at a given time. The complexity of the resource management depends on the system size and performing an exhaustive search of the optimal schedule swiftly becomes unfeasible as both number of users and spot beams increase. For this reason efficient scheduling algorithms need to be designed \cite{BoBeRo11}.

Based on the load of the return-link, it is possible to reserve additional resources in order to reduce the number of users transmitting at the same time. This strategy can help improving the Quality of Service (QoS) perceived at higher layers. However the utilization of additional resources implies a loss of efficiency.

In this paper we assume universal frequency reuse and we focus more particularly on the resource management performed by the gateway on the return-link. We consider clear sky conditions and investigate the impact of the number of users and spot beams on the scheduling performance,  evaluated in terms of minimum achievable per-user rates. For this purpose a scheduling algorithm based on a bipartite graph approach is considered. The utilization of additional resources is referred as free slot assignment (FSA) in the following and will help guaranteeing minimum per-user rates at the expense of efficiency. Therefore a specific algorithm needs to be defined in order to avoid wasting capacity.

The paper is organized as follows. Section~\ref{s:scenarion_and_bg} describes the scenario considered in this study and details the bipartite graph scheduling algorithm. Section~\ref{s:impact_scheduling_size} shows the impact of the number of users and beams on the minimum per-user rates achieved in the system. The principle of FSA and its performance are addressed in Section~\ref{s:fsa}. The conclusions are finally drawn in Section~\ref{sec:Conclusions}.

\section{Scenario and Bipartite Graph Scheduling}\label{s:scenarion_and_bg}

\begin{figure}[h]
  \begin{center}
  \includegraphics[width=0.35\textwidth]{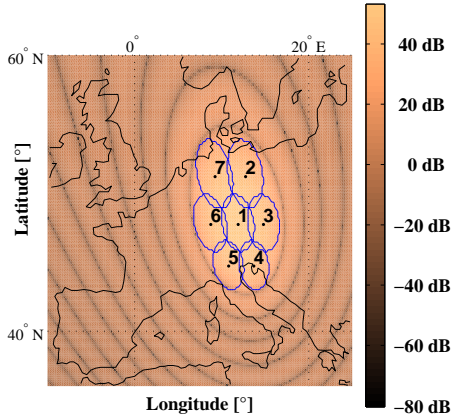}
  \caption{Multi-beam scenario considered with $B=7$ beams. The radiation of the beam in the middle is shown.}
  \label{fig:K1_7x7_radiation_1}
 \end{center}
\vspace{-15pt}
\end{figure}

The multi-beam satellite system considered is the same as in \cite{BBRJ11}. The satellite is geo-stationary with longitude $19.2^\circ$ East and has a transparent payload. Its topology describes a star network and the service area covers Europe with 96 spot beams, simply referred as beams in the following. In this paper we consider a subset of $B=7$ beams as depicted in (Fig.~\ref{fig:K1_7x7_radiation_1}). The beam in the middle\footnote{Beam center coordinates: latitude $48.75^\circ$ North and longitude $11.9^\circ$ East} covers the city of Munich and is surrounded by six other beams directly adjacent and using the same frequency.

The satellite terminals (users) are randomly and uniformly distributed in the different beams in such a way that the number of users per beam, denoted as $M$, is identical in all $B$ beams. We assume slotted time in this paper with one single user assigned in each beam and slot. In a slot, the association of users located in $B$ different beams is called combination. The set of $M$ combinations of users constitute the scheduling output, also named allocation.  We define with $N_\text{users,sched}$ the number of users taken into account for determining an allocation and also referred as the scheduling size:
\vspace{-3pt}
\begin{equation}
N_\text{users,sched}=M\,B
\vspace{-3pt}
\end{equation}
Therefore both parameters influence the scheduling: $B$ represents the system size and $M$ the scheduling depth. The number of possible allocations, is:
\vspace{-3pt}
\begin{equation}\label{eq:nb_allocations}
N_\text{alloc}=\left(M!\right)^{B-1}
\vspace{-3pt}
\end{equation}
The order of the slot is considered as not relevant as seen in (\ref{eq:nb_allocations}) ($B-1$ term). As soon as $M>2$ redundancy appears in the list of possible allocations. Actually different allocations can present the same combination of users in some slots.

\begin{figure}[h]
  \begin{center}
  \includegraphics[height=5cm]{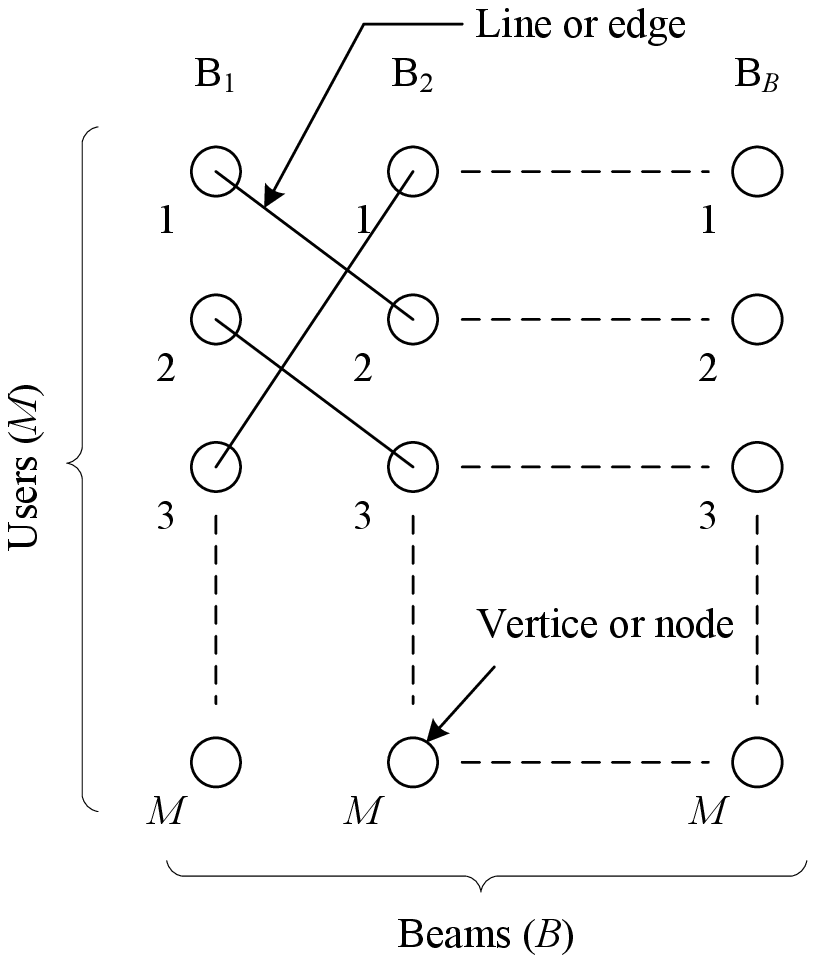}
  \caption{Terms definition in the bipartite graph approach.}
  \label{fig:scheduling-graph_terms}
 \end{center}
\vspace{-22pt}
\end{figure}

A bipartite graph approach \cite{CoLeRi90} was applied in order to avoid redundancy and operates directly at the combination level. The nodes (or vertices) correspond to the transmitters in our system, more particularly the index of the nodes refers to the index of the user in the covering beam. An edge (or line) represents the scheduling of a user in a given beam with a user in another beam. An edge is thus the combination of two users. Fig.~\ref{fig:scheduling-graph_terms} summarizes the different terms. We additionally define the term "path" as a combination of users scheduled at the same time in all $B$ beams. An allocation is therefore the association of $M$ paths in this bipartite graph approach.

The number of possible edges with two beams, i.e. the number of possible combination of users is $N_{\text{edges},B=2}=M^2$. The number of possible paths, i.e. the number of possible combination of users in all beams, is therefore:
\vspace{-3pt}
\begin{equation}
N_\text{paths}=M^B
\vspace{-3pt}
\end{equation}

\begin{figure}[h]
  \begin{center}
  \includegraphics[width=9cm]{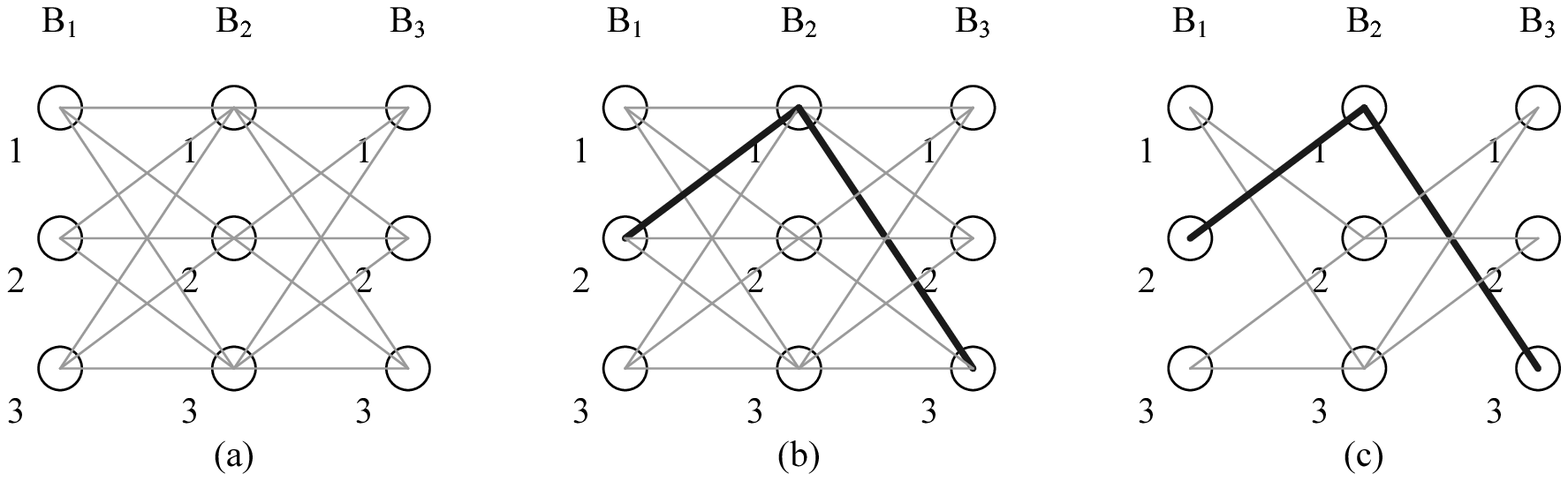}
  \caption{Example of bipartite graph.}
  \label{fig:scheduling-graph_example}
 \end{center}
\vspace{-12pt}
\end{figure}

The principle for determining a schedule is to evaluate the achievable per-user rates for each path and to select $M$ paths, depending on the adopted strategy. For each selection of a path the nodes involved in this path have to be pruned since users cannot be scheduled more than once. An example is given by Fig.~\ref{fig:scheduling-graph_example} where $M=3$ and $B=3$. In the left graph (a) all possible edges are shown. In (b) a path is selected: user \#2 in beam 1, user \#1 in beam 2 and user \#3 in beam 3 are scheduled at the same time. Finally in (c) the edges involving nodes of the previous selected path are pruned, reducing the number of remaining combinations.

On one side an exhaustive search among all possible allocations, denoted as ES, results in a search space given by (\ref{eq:nb_allocations}). On the other side an exhaustive search realized with the bipartite graph approach (BG) operates at the combination level. Since the rates are computed on a slot basis, the number of evaluations to be performed is:
\begin{equation}
N_\text{eval,ES}=M\,\left(M!\right)^{B-1}
\hspace{0.7cm}
N_\text{eval,BG}=M^B
\end{equation}
For this reason the bipartite graph approach permits to reduce the number of evaluations as soon as $M>2$.\\

Each generation of users randomly placed in the coverage area of the beams corresponds to a channel matrix $\boldsymbol{H}^{(c)}$ where $c$ is the channel matrix index and each channel matrix is of size $N_\text{Rx}\times N_\text{Tx}$ ($N_\text{Tx}$: number of transmitters, $N_\text{Rx}$: number of receivers. In our case $N_\text{Tx}=N_\text{Rx}=B$). As already mentioned scheduling consists into combining different users at the same time. Therefore an allocation will result into the association of columns issued from different channel matrices. For instance with a $3\times{3}$ system and two users per beam (i.e. $B=3$ and $M=2$), we have two initial channel matrices: $\boldsymbol{H}^{(1)}$ and $\boldsymbol{H}^{(2)}$. $\boldsymbol{H}^{(1)}$ corresponds to the channel matrix of the users with index "1" and $\boldsymbol{H}^{(2)}$ to the channel matrix of the users with index "2". A possible allocation is:
\begin{equation}\nonumber
\boldsymbol{H}^{(1)'}=\left[
\boldsymbol{h}^{(2)}_1\,\,\boldsymbol{h}^{(1)}_2\,\,
\boldsymbol{h}^{(2)}_3
\right]
\hspace{1cm}
\boldsymbol{H}^{(2)'}=\left[
\boldsymbol{h}^{(1)}_1\,\,\boldsymbol{h}^{(2)}_2\,\,
\boldsymbol{h}^{(1)}_3
\right]
\end{equation}
where $\boldsymbol{h}^{(c)}_j$ represents the $j$-th column vector of the $c$-th channel matrix. $\boldsymbol{H}^{(c)'}$ corresponds to the new channel matrix issued from the combination of users in the $c$-th slot. In other words scheduling consists into swapping one or more
columns of different channel matrices.

\begin{figure}[h]
  \begin{center}
  \includegraphics[width=9cm]{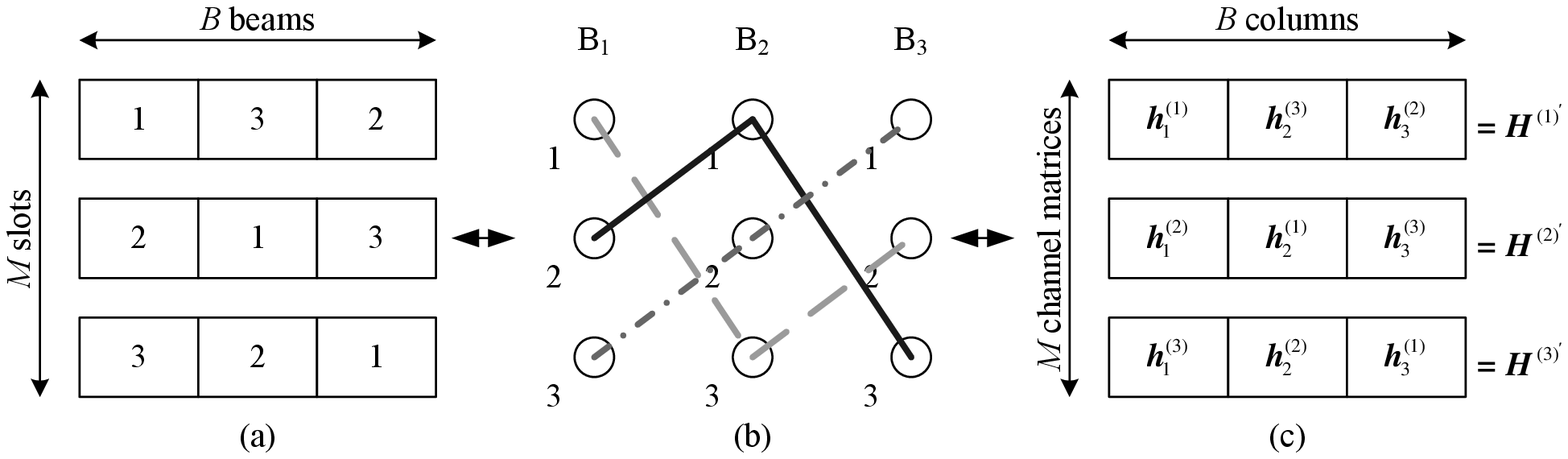}
  \caption{Relation between allocation, bipartite graph and channel matrices.}
  \label{fig:alloc-bg-matrices}
 \end{center}
\vspace{-15pt}
\end{figure}

In fig.~\ref{fig:alloc-bg-matrices} the relation between an example of allocation (a), the corresponding bipartite graph (b) and the resulting channel matrices (c) is depicted with $B=3$ and $M=3$.

The achievable rate of user $j$, initially belonging to $\boldsymbol{H}^{(c)}$, is a function of the new channel matrix in which it is scheduled and also the order in which it will be decoded when performing SIC. Since the target of this study is the maximization of the minimum per-user rate through the jointly use of these two elements, concerning SIC the optimal ordering defined in \cite{BBRJ11} is used in the following. The signal-to-interference plus noise ratio (SINR) experienced by this user is:
\begin{equation}\label{eq:sinrj_mmse-sic}
\text{SINR}_{j,\text{MMSE-SIC}}^{(c)'}=
\boldsymbol{h}_{j}^{(c)\,\text{H}}\left(
\boldsymbol{C_\eta}+\sum_{i'>{j'}}\boldsymbol{h}^{(c)'}_{i'}\,
\boldsymbol{h}_{i'}^{(c)'\,\text{H}}\right)^{-1}\,
\boldsymbol{h}^{(c)}_{j}
\end{equation}
where $i'$ and $j'$ are the new column indexes of users $i\neq j$ and user $j$ after ordering (please note that the content of the column vector does not change, therefore $\boldsymbol{h}^{(c)}_j=\boldsymbol{h}^{(c)'}_{j'}$).  $\boldsymbol{A}^\text{H}$ denotes the Hermitian of the matrix $\boldsymbol{A}$, and $\boldsymbol{C_\eta}$ represents the covariance of the noise vector $\boldsymbol{\eta}\sim\mathcal{C}\,\mathcal{N}\left(0,N_0\,\boldsymbol{I}\right)$. As seen the equivalent SINR in (\ref{eq:sinrj_mmse-sic}) assumes MMSE filtering together with SIC. The achievable rate of each user $j$, expressed in bits/s/Hz, can be simply evaluated as \cite{Sha49}:
\begin{equation}
R_{j,\text{MMSE-SIC}}^{(c)'}=
\log_2\left(1+\text{SINR}_{j,\text{MMSE-SIC}}^{(c)'}\right)
\end{equation}

In order to make the selection of the $M$ paths we consider a minimum deletion algorithm which operates as follows. The paths are sorted according to the minimum per-user rate achieved in each slot. One by one the path with the lowest minimum per-user rate is removed from the list. The deletion of a path may be performed provided that with the remaining nodes there is still a solution to the scheduling problem. If no solution remains then this path is part of the final allocation. The corresponding nodes are pruned and the process ends when the $M$ paths are found. The principle of this algorithm is derived from \cite{BaCaZo08} and was adapted and exploited in \cite{BoBeRo11}.

\section{Impact of the Scheduling Size}\label{s:impact_scheduling_size}

Based on thousands of channel matrices the impact of the scheduling size on the minimum per-user rate was investigated. The number of channel matrices is denoted as $N_\text{ch}$, therefore the total number of users in the system is:
\begin{equation}
N_\text{users,total}=N_\text{ch}\,B
\end{equation}
The number of schedules to be determined is:
\begin{equation}
N_\text{sched}=
\bigg\lceil\frac{N_\text{users,total}}{N_\text{users,sched}}\bigg\rceil
=\bigg\lceil\frac{N_\text{ch}}{M}\bigg\rceil
\end{equation}
where $\big\lceil\cdot\big\rceil$ represents the ceiling operator.

In this study we consider a signal-to-noise ratio (SNR) of 15~dB such that the communication is mainly interference-limited. On the one hand increasing the number of beams $B$ is equivalent to add more interferers in the system. On the other hand, with fixed $B$, having more users $M$ in each beam extends the degree of freedom for scheduling. For investigating the impact of $B$ on scheduling, the bipartite graph approach presented in the previous section is used. For $B=7$ weakly interfering users can be grouped in fictitious colors and be randomly scheduled reducing the number of paths to be drawn \cite{BoBeRo11}. Identical performance in terms of max-min rate can be reached in this way while reducing the processing complexity.

The beams are numbered such that $B_i=i$ with $i=1\,...\,B$ (the mapping of $i$ is shown by Fig.~\ref{fig:K1_7x7_radiation_1}). The fairness between the users is measured based on the achievable rates obtained in each slot, i.e. with each $\boldsymbol{H}^{(c)'}$, and using the Jain's fairness index \cite{JaHaCh84}. We additionally denote the non-availability or outage probability for a given minimum rate $r$ as:
\begin{equation}\label{eq:out_prob}
p_{r}=P\{r_\text{min}\leqslant{r}\}
\end{equation}
In the rest of this paper the cumulative distribution function (CDF) of the minimum per-user rates will be interpreted using (\ref{eq:out_prob}).

\vspace{-10pt}
\begin{figure}[h]
  \begin{center}
  \includegraphics[width=0.5\textwidth]{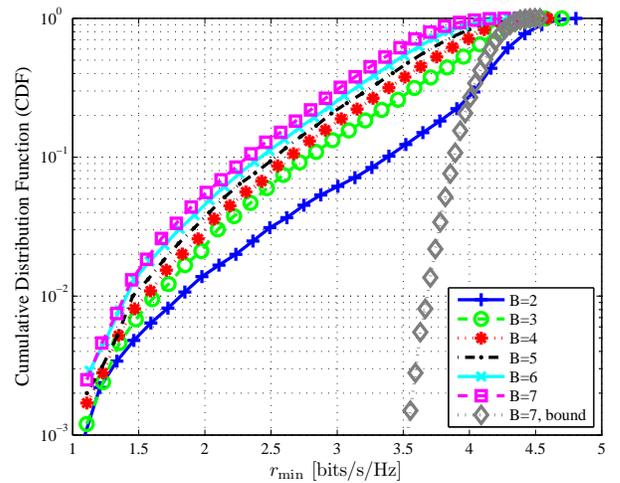}
  \caption{Impact of the number of beams on the minimum per-user rates.}
  \label{fig:analyse_impact_sched_size-B_M1-C}
 \end{center}
\vspace{-15pt}
\end{figure}

Fig.~\ref{fig:analyse_impact_sched_size-B_M1-C} shows the impact of $B$ on the CDF of the minimum per-user rates when having $M=1$ (i.e. no scheduling). On one side, as expected, we notice a decrease of performance while $B$ gets higher. For a given $p_r$, the corresponding minimum guaranteed rate $r$ decreases. The outage probability $p_{r=3}$ ranges from about 0.06 to 0.31 respectively for $B=2$ and $B=7$. From $B=2$ to $B=3$ there is a remarkable jump since the particular disposition of the beams rises considerably the probability of having interferers. On the other side the decrease of performance is relatively moderate as $B$ grows since the gain offered by the MIMO system (multiple receive antennas) is balanced with the increasing level of CCI. It has to be noticed that the lowest minimum per-user rate remains the same and is almost 1.1 bits/s/Hz even though the probability to occur decreases with smaller values of $B$. Fig.~\ref{fig:analyse_impact_sched_size-B_M1-C} also shows the per-user rate bound for $B=7$ which is simply the sum rate divided by the number of users. Therefore the mean per-user rate represents the upper-bound in which all users experience the same rate.

\begin{figure}[h]
  \begin{center}
  \includegraphics[width=0.5\textwidth]{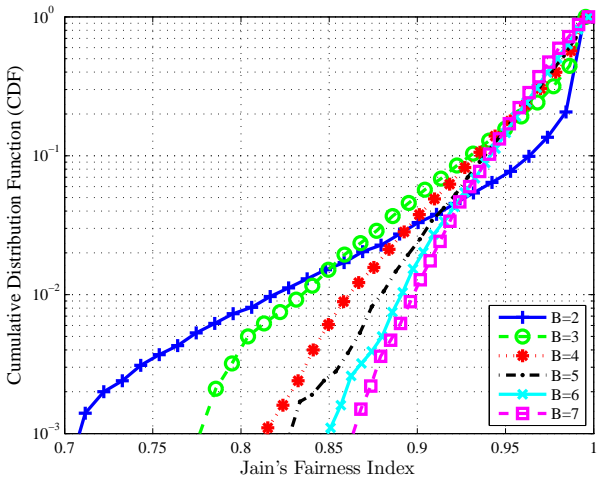}
  \caption{Impact of the number of beams on the Jain's fairness index.}
  \label{fig:analyse_impact_sched_size-B_M1-J}
 \end{center}
\vspace{-15pt}
\end{figure}

On the contrary the impact of $B$ on the Jain's fairness index is such that it increases for larger $B$ (Fig.~\ref{fig:analyse_impact_sched_size-B_M1-J}). Actually the users are more likely to have different SINRs and this improves the performance of SIC with optimal ordering. The lowest fairness index ranges from 71\% with $B=2$ to 86\% with $B=7$.

\vspace{-10pt}

\begin{figure}
  \begin{center}
  \includegraphics[width=0.5\textwidth]{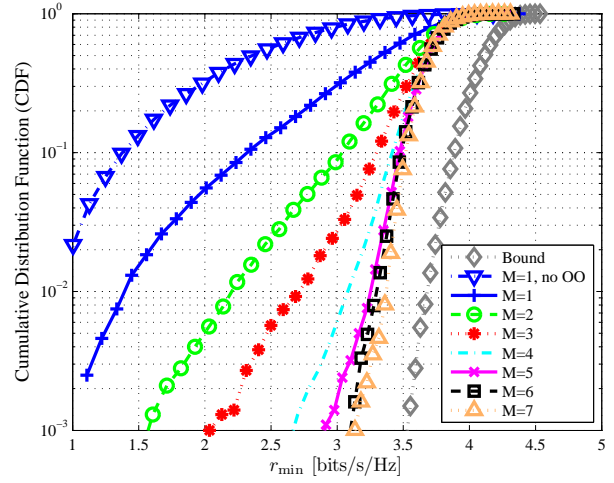}
  \caption{Impact of the number of users on the minimum per-user rates.}
  \label{fig:analyse_impact_sched_size-B7_M-C}
 \end{center}
\vspace{-15pt}
\end{figure}

The next step consists into fixing the number of beams such that $B=7$.  As already mentioned the degree of freedom for scheduling increases with the number of users per beam $M$. Actually the probability of finding a less interfered combination of users on ground grows with $M$. It is thus foreseen to observe an improvement of performance. The purpose is to get as close as possible to the upper-bound which represents the ideal case of perfect balance between users' rates with a consequently unitary Jain's index. In fig.~\ref{fig:analyse_impact_sched_size-B7_M-C} the CDF shows that for a given $p_r$ the minimum per-user rates are clearly improved as the scheduling size gets higher. However the gain decreases with $M$ and with $p_r=0.01$ five users per beam already saturates. Actually, due to the number of interferers present in the system, it is difficult to reach the upper-bound. Therefore for such low probability having $M$ greater than 5 will not bring additional performance advantages. In terms of performance, with $p_r=0.1$ the minimum per-user rate ranges from 1.4 to 3.5 bits/s/Hz depending on the ordering and $M$.

\begin{figure}
  \begin{center}
  \includegraphics[width=0.5\textwidth]{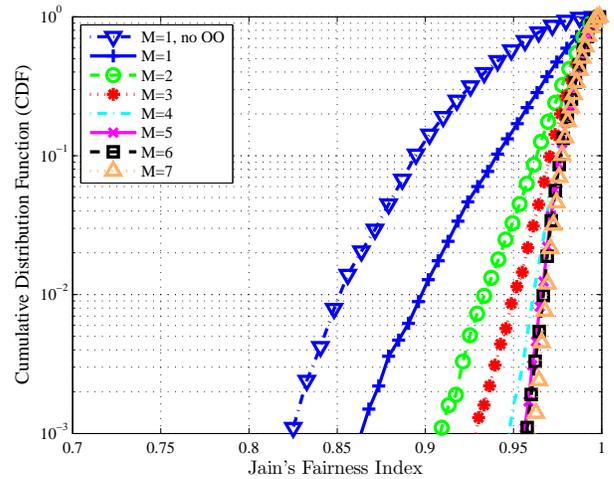}
  \caption{Impact of the number of users on the Jain's fairness index.}
  \label{fig:analyse_impact_sched_size-B7_M-J}
 \end{center}
\vspace{-25pt}
\end{figure}

Looking at Fig.~\ref{fig:analyse_impact_sched_size-B7_M-J}, the fairness is also dramatically improved as soon as scheduling is performed. However, as before, we observe that starting from $M=5$ the gain is almost negligible. With this number of users per beam we can already provide a Jain's fairness index greater than 95\%. As a conclusion scheduling with $M=5$ provides close to optimal performance in this $7\times 7$ satellite MIMO system.

Considering uniformly distributed users within the wide beam footprint of modern satellite systems, high SINRs unbalances are likely to occur in the return-link, even in clear sky conditions. The resource allocation for $M$ users can be optimized by the gateway by means of scheduling algorithms. Furthermore, the jointly use of SIC ordering contributes in lowering the scheduling depth required to reach the system target.

\section{Free Slot Assignment}\label{s:fsa}

The satellite operator may be interested in increasing the minimum rates and also the level of fairness, and this can be done lowering CCI. To do so a scheduling mechanism needs to be designed, triggered by the statistics obtained so far.

One way of reducing CCI is to avoid having necessarily $B$ users transmitting at the same time. This strategy is referred as free slot assignment (FSA) in this paper and in the following one single free slot per schedule is assumed. Therefore $M$ users are allocated in $M+1$ slots. It can easily be shown that the number of possible allocations, when avoiding the presence of empty slots (unused), is increased to:
\begin{equation}
N_{\text{alloc,FSA}}=
\left((M+1)!\right)^{B-1}-\left(M!\right)^{B-1}
\end{equation}

The minimum achievable rate to be guaranteed to the users is referred as $r_\text{th}$ in the following. The utilization of an additional slot may not be necessary if $r_\text{th}$ can be already achieved with or without scheduling, i.e. with $M>1$ or $M=1$, respectively. This permits to avoid wasting resources uselessly. An algorithm was designed for this purpose and operates as follows.

The channel matrices can be seen as the users joining the system and requesting resources on the RL. They are considered one by one until all of them are processed (i.e. all users got resources). The number of users per beam, which influences the scheduling process, is initially set to $M=1$. The minimum per-user rate is computed on a slot basis and is compared to the minimum per-user rate requested $r_\text{th}$. If the result is lower, the number of users per beam is increased by one, i.e. $M=2$ and scheduling is performed with the current channel matrix and the next channel available. The process is repeated until the requested minimum per-user rate can be reached or until $M$ exceeds the FSA threshold, denoted as $M_\text{FSA}$. In the latter case FSA is enabled and a last try is performed using $M_\text{FSA}+1$ slots.

It may happen that even with a free slot assignment the minimum per-user rate requested cannot be reached. In this case the last step of the scheduling process can be ignored and the process can continue with the next channel matrix available. However, the scheduling output with $M_\text{FSA}+1$ slots will provide the highest minimum per-user rate even if the constraint is not satisfied, suggesting to keep the final scheduling.

As before simulations are performed at SNR 15~dB and run over thousands channel matrices. The values of $M_\text{FSA}$ and $r_\text{th}$ are selected in order to evaluate the FSA performance with $p_{r_\text{th},M}\simeq{0.1}$. For instance according to Fig.~\ref{fig:analyse_impact_sched_size-B7_M-C} with $M=2$ this probability is reached when $r_\text{th}\simeq{3}$~bits/s/Hz. An estimation of the frequency at which a free slot gets assigned to the users is:
\vspace{-5pt}
\begin{equation}\label{eq:f_fsa}
f_\text{FSA}=p_{r_\text{th},M}\times\frac{1}{M_\text{FSA}}
\vspace{-3pt}
\end{equation}
One could then expect with $M_\text{FSA}=2$ to have about 5\% of FSA use. However the $f_\text{FSA}$ value issued from simulations differs from (\ref{eq:f_fsa}) since the parameter $M$ changes together with the association of channel matrices for each schedule. Moreover in (\ref{eq:f_fsa}) it is assumed that the outage only happens once per schedule, which is not necessarily the case in reality.

Table~\ref{table:stats_nofsa_scheduling_B7} and Table~\ref{table:stats_fsa_scheduling_B7} summarize the performance obtained without FSA (benchmark) and with FSA, respectively. "Slots without sched." gives the percentage of slots for which scheduling was not necessary (i.e. $M=1$). "Slots with $M=x$" and "Slots availability" indicate respectively the percentage of slots for which the scheduling size is $M=x$ and the percentage of slots for which $r_\text{th}$ could be guaranteed. The "Efficiency" quantifies the loss of resources due to FSA. Finally "FSA use" represents the percentage of additional free slots. What is remarkable, as shown in Table~\ref{table:stats_fsa_scheduling_B7}, is the capacity of the algorithm to detect when scheduling is not mandatory or when a lower $M$ reaches the targeted rate. This reduces considerably the complexity and therefore the computation time for optimizing the user schedules. It has also to be remarked that increasing the degree of freedom $M_\text{FSA}$, we are able to guarantee higher $r_\text{th}$ constraints while keeping the efficiency almost constant.

\begin{table}
 \begin{center}
  \caption{Scheduling performance without FSA.}
  \begin{tabular}{|l|c|c|c|}
  \hline
  $M$ [users per beam] & $2$ & $3$ & $4$\\
  \hline
  Service rate req. $r_\text{th}$ [bits/s/Hz] & $3$ & $3.3$ & $3.41$\\
  \hline
  Slots with $M=2$ [\%] & $100.0$ & - & -\\
  \hline
  Slots with $M=3$ [\%] & - & $100.0$ & -\\
  \hline
  Slots with $M=4$ [\%] & - & - & $100.0$\\
  \hline
  Slots availability [\%] & $91.08$ & $89.63$ & $90.11$\\
  \hline
  Efficiency [\%] & $100.0$ & $100.0$ & $100.0$\\
  \hline
  \end{tabular}
  \label{table:stats_nofsa_scheduling_B7}
 \end{center}
\vspace{-10pt}
\end{table}

\begin{table}
 \begin{center}
  \caption{Scheduling performance with FSA.}
  \begin{tabular}{|l|c|c|c|}
  \hline
  $M_\text{FSA}$ [users per beam] & $2$ & $3$ & $4$\\
  \hline
  Service rate req. $r_\text{th}$ [bits/s/Hz] & $3$ & $3.3$ & $3.41$\\
  \hline
  Slots without sched. [\%] & $56.46$ & $34.04$ & $25.18$\\
  \hline
  Slots with $M=2$ [\%] & $36.98$ & $29.24$ & $23.80$\\
  \hline
  Slots with $M=3$ [\%] & $6.56^1$ & $31.39$ & $21.18$\\
  \hline
  Slots with $M=4$ [\%] & - & $5.33^1$ & $26.06$\\
  \hline
  Slots with $M=5$ [\%] & - & - & $3.78^1$\\
  \hline
  Slots availability [\%] & $99.46$ & $98.89$ & $99.20$\\
  \hline
  Efficiency [\%] & $93.84$ & $94.94$ & $96.36$\\
  \hline
  \end{tabular}
  \label{table:stats_fsa_scheduling_B7}
 \end{center}
\vspace{-5pt}
\hspace{30pt}$^1$ \textit{also equivalent to the percentage of FSA use.}
\vspace{-15pt}
\end{table}

\begin{figure}
  \begin{center}
  \includegraphics[width=0.5\textwidth]{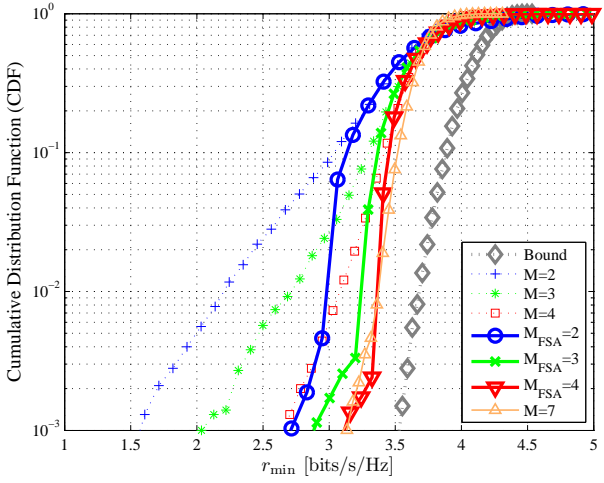}
  \caption{Impact of FSA on the minimum per-user rates.}
  \label{fig:analyse_fsa_sched-B7_M-C}
 \end{center}
\vspace{-15pt}
\end{figure}

\begin{figure}
  \begin{center}
  \includegraphics[width=0.5\textwidth]{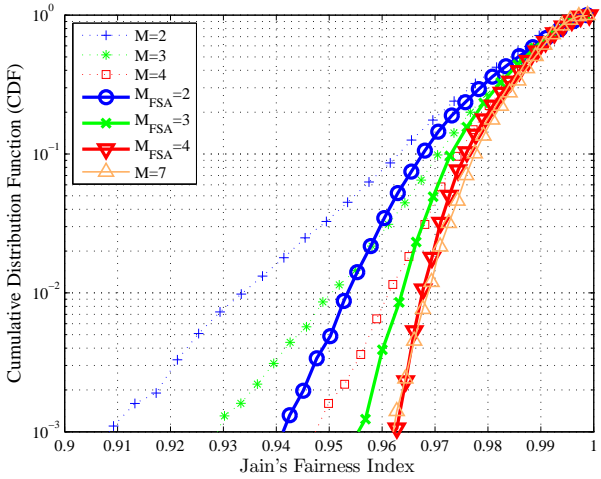}
  \caption{Impact of FSA on the Jain's fairness index.}
  \label{fig:analyse_fsa_sched-B7_M-J}
 \end{center}
\vspace{-15pt}
\end{figure}

Fig.~\ref{fig:analyse_fsa_sched-B7_M-C} shows the corresponding CDFs of the minimum per-user rates. As expected we clearly see that FSA increases radically the lowest minimum per-user rates. The proposed algorithm even performs as well as scheduling without FSA and with larger $M$ ($M_\text{FSA}=4$ achieves almost identical performance as $M=7$). This trend can also been observed in Fig.~\ref{fig:analyse_fsa_sched-B7_M-J} when analyzing the fairness between users and is actually a consequence of the results previously obtained. The lowest Jain's fairness index is therefore improved from around 82\% without scheduling and without optimal ordering (Fig.~\ref{fig:analyse_impact_sched_size-B7_M-J}) to more than 96\% with FSA and $M_\text{FSA}=4$.

The complexity of this algorithm cannot be easily predicted since $M$ and the use of a free slot are conditioned by $r_\text{th}$ and the level of CCI. According to \cite{BoBeRo11}, the complexity of a schedule with this algorithm, without FSA, and scaled to a slot basis, is proportional to:
\vspace{-5pt}
\begin{equation}
\alpha_M=M^4\,\left(C-1\right)\,\frac{1}{M}=2\,M^3
\end{equation}
since $C=3$. The average complexity of a schedule with the proposed FSA algorithm is proportional to:
\vspace{-5pt}
\begin{equation}\label{eq:fsa_comp}
\alpha_{M_\text{FSA}}=
\sum_{i=1}^{M_\text{FSA}}\,\frac{f_i}{i}\,\sum_{j=1}^i\,2\,j^4+
\frac{f_{M_{\text{FSA}}+1}}{M_\text{FSA}}
\,\sum_{j=1}^{M_\text{FSA}+1}\,2\,j^4
\vspace{-3pt}
\end{equation}
where $f_i$ is the frequency at which slots with $M=i$ are used, therefore $f_{M_{\text{FSA}}+1}=f_\text{FSA}$. The left term in (\ref{eq:fsa_comp}) represents the complexity without the presence of an additional free slot, the inner sum is related to the tries which are performed while increasing $M$. The right term relates to the complexity when FSA is enabled. The gain in complexity reduction increases with $M_\text{FSA}$ since the probability of finding an intermediate $M\leqslant M_\text{FSA}$ fulfilling the requirement gets higher (Table~\ref{table:gain_complexity}).

\begin{table}[h]
\vspace{-5pt}
 \begin{center}
  \caption{Complexity analysis: proposed FSA algorithm vs no FSA.}
  \begin{tabular}{|l|c|c|c|}
  \hline
  $M$ or $M_\text{FSA}$ [users per beam] & $2$ & $3$ & $4$\\
  \hline
  Service rate req. $r_\text{th}$ [bits/s/Hz] & $3$ & $3.3$ & $3.41$\\
  \hline
  Gain wrt. no FSA [\%] & $13.47$ & $28.26$ & $35.14^1$\\
  \hline
  \end{tabular}
  \label{table:gain_complexity}
 \end{center}
\vspace{-5pt}
\hspace{20pt}$^1$ \textit{$87.90\%$ gain with respect to $M=7$.}
\end{table}

\vspace{30pt}

\section{Conclusions}\label{sec:Conclusions}
\vspace{-2pt}

This paper investigated the impact of resource management in the return-link of an interference-limited multi-beam satellite system, with full frequency reuse. The satellite multiple-input multiple-output system has been described in terms of per-user achievable rates assuming clear sky conditions and perfect channel state information. In order to maximize the minimum per-user rate, the MIMO capacity achieving architecture, with optimal successive interference cancellation ordering, has been assumed to quantify the impact of the scheduling size. The impact of the latter was double. On the one hand, as expected, the level of co-channel interference increased with the system size and the gain of multiple receiving antennas balanced the decrease of performance. On the other hand while increasing the number of users per schedule, a convergence to the upper-bound and a saturation of the performance have been observed. With the users population uniformly distributed into the spot beams, a reduced scheduling depth permitted to reach close to optimal performance in terms of minimum per-user rates but also in terms of fairness. Finally, tuning the system load, it was possible to reserve part of the resources in order to guarantee a minimum rate with a negligible failure probability. Based on the reservation of a free slot, the proposed algorithm provided a good trade-off between user satisfaction and efficiency-loss while considerably saving on scheduling complexity.

\vspace{-3pt}
\section*{Acknowledgment}
\vspace{-2pt}

The authors would like to thank Dr. Francesco Rossetto for the useful discussions and hints helping in preparing this paper.

\end{document}